\documentclass{mem}  
\usepackage{natbib}
\usepackage{txfonts}
\usepackage{balance}
\usepackage{graphicx}
\usepackage[a4paper]{hyperref}

\begin{document}

\title{Spectropolarimetric inversions of the He I 10830 \AA\ multiplet in an Active Region filament}

   \subtitle{}

\author{
C. \,Kuckein\inst{1,2} 
\and R. \,Centeno\inst{3}
\and V. \,Mart{\'i}nez Pillet\inst{1,2}
          }

  \offprints{C. Kuckein}

\institute{
Instituto de Astrof{\'i}sica de Canarias,
C/ V{\'i}a L{\'actea} s/n, 
E-38205 La Laguna, Tenerife, Spain
\email{ckuckein@iac.es}
\and Departamento de Astrof{\'i}sica, Universidad de La Laguna, E-38206 La Laguna, Tenerife, Spain
\and High Altitude Observatory (NCAR), Boulder, CO 80301, USA
}

\authorrunning{Kuckein et al.}

\titlerunning{Inversions of the He I 10830 \AA\ multiplet in an AR filament}

\abstract{
Full Stokes spectropolarimetric data (in the 10830\,\AA\ region) of an active
region filament were obtained in July 2005 using the Tenerife Infrared
Polarimeter instrument. The polarization profiles in the filament show
Zeeman-like signatures. Milne-Eddington inversions were performed to infer the
chromospheric magnetic field, inclination, azimuth, velocity and Doppler width
from the \ion{He}{I} 10830\,\AA\ multiplet. Field strengths of the order of
600-800\,G were found in the filament. Strong transverse fields at
chromospheric levels were detected near the polarity inversion line. To our
knowledge, these are the highest field strengths reliably measured in these
structures. Our findings suggest the possible presence of a flux rope. 
  
}

\maketitle{}

\section{Introduction}
Filaments can be seen as long and dark structures in H$\alpha$ observations on
the solar disk and are formed of denser and cooler plasma than their
surroundings. We can distinguish between two types of filaments: active region
(AR) and quiescent filaments. They are located along polarity inversion lines
(PILs) and their exact height is not easily inferred from on-disk data. From observations
and three-dimensional models it has been showed that AR filaments are stronger
and lie lower in the atmosphere than their quiescent counterparts
\citep[see][and references therein]{ck-AuDe03}. 

Partly due to the limited availability of spectropolarimetric data,
there are very few studies of magnetic field strengths in AR filaments. \citet{ck-wiehr91} inferred the
longitudinal magnetic field in an AR prominence using the Stokes $V$ spectra of
the \ion{Ca}{II} $8542$~\AA. Values between 75 and 180\,G were found. Furthermore, by analyzing the full Stokes vector of the \ion{He}{I} $10830$\,\AA\ of a multicomponent flaring active
region, \citet{ck-sasso07} have inferred magnetic field strengths around 380\,G.

\begin{figure*}
\includegraphics[width=0.48\hsize]{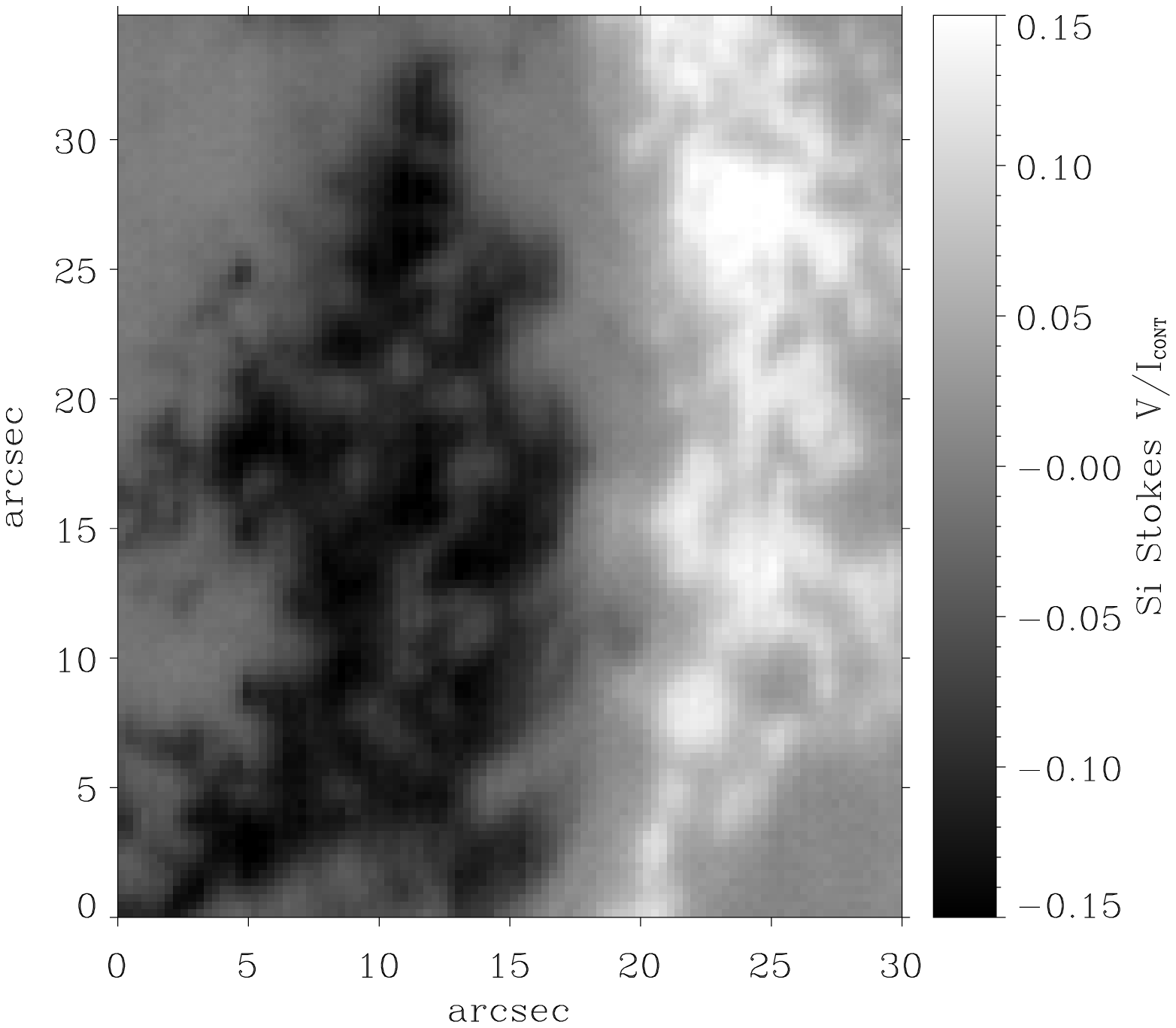}
\includegraphics[width=0.48\hsize]{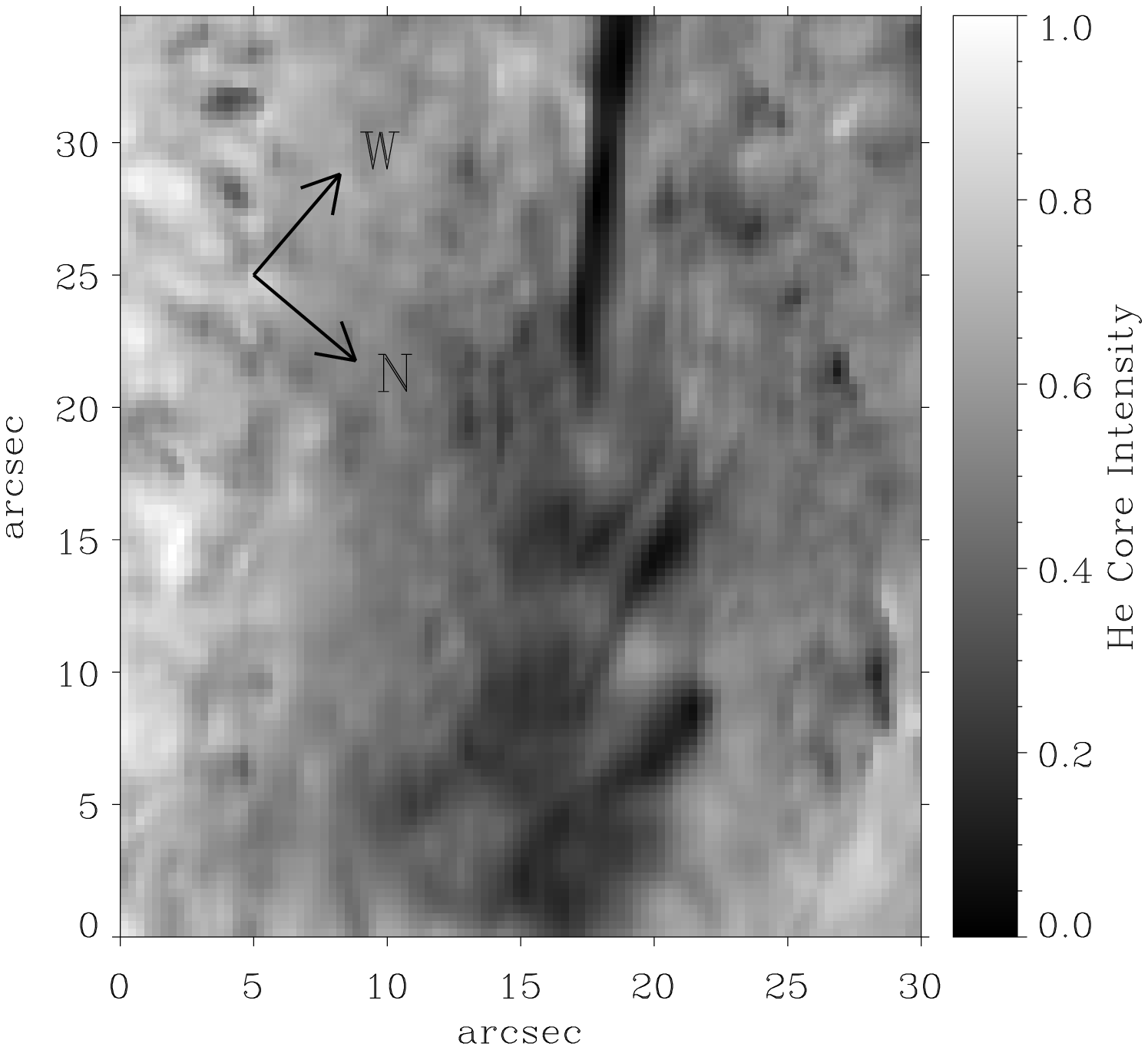}
\caption{
\footnotesize \textit{Left}: \ion{Si}{I} Stokes $V$ map normalized to the continuum intensity. \textit{Right}: \ion{He}{I} ``red'' core intensity map reconstructed from the slit scan positions. The arrows indicate Solar West and North. }
\label{kuckein-fig:Hecoremap}
\end{figure*}

This paper will focus on the full Stokes inversion of \ion{He}{I} $10830$\,\AA\
multiplet measured in an AR filament near the solar disk center.
Section 2 begins by laying out the observations, section 3 describes the
spectropolarimetric inversions and results. The last section assesses the
discussion.

\section{Observations}
The analyzed data were acquired on the 5th of July, 2005, with the Tenerife Infrared Polarimeter \citep[TIP-II,][]{ck-tip2} at the German Vacuum Tower Telescope (VTT, Tenerife, Spain). The active region NOAA 10781 was localized near the disk center at N13-W29 around $\mu = 0.92$. The filament was lying above a compact active region neutral line. 

Various spatial scans, using in addition the adaptive optics system KAOS \citep{ck-kaos03}, were taken along the AR with the slit parallel to the polarity inversion line, covering an area of $35\arcsec \times 30\arcsec$. Moreover, one time series was acquired with the slit fixed at the PIL. The spectral range covered the chromospheric \ion{He}{I} $10830$~\AA\ multiplet, the photospheric \ion{Si}{I} $10827.1$~\AA\ line and at least one telluric line with a spectral sampling of $\sim 11.1$\,m\AA\,px$^{-1}$. The data reduction, as well as the polarimetric calibration \citep{ck-collados99,ck-collados03}, were performed for all the data-sets. In order to improve the signal-to-noise ratio a 3px binning in the spectral domain and a 6px binning in the spatial direction were applied to the data. For a more detailed description of the observations see \citet{ck-kuck09}.

\subsection{Description}
The \ion{He}{I} ``red'' component at $\sim 10830.3$\AA\ slit-reconstructed
image in the \textit{right} panel of Figure \ref{kuckein-fig:Hecoremap} shows a
dense concentration of plasma above the PIL. Highly twisted structures can be
identified in the lower part of the map. On the contrary, in the upper part
these dark structures seem to align with the PIL. These findings could suggest
the presence of a flux rope. One question that needs to be asked, however, is
whether we are seeing the dips or the tops of that flux rope. An exhaustive study
of the vector magnetic field must be carried out to obtain this information.
This is future work that will be presented in our next paper. 

The Stokes profiles of \ion{He}{I} $10830$\AA\ obtained in this campaign
\citep[see Fig. 1 of][]{ck-kuck09} indicate a strong predominance of
Zeeman-like signatures in the AR filament,  characteristically with three-lobe
profiles in Stokes $Q$ and $U$. Observations of quiescent filaments using the
same \ion{He}{I} multiplet have shown that the polarization is dominated by
atomic level polarization and its modification through the Hanle effect
\citep[see][]{ck-truji02}. 

\section{Spectropolarimetric inversions}
Three methods were used to invert the Stokes profiles: magnetograph analysis, Milne-Eddington (ME) inversions and principal component analysis (PCA), the latter taking into account the physics of atomic level polarization and the Hanle effect \citep[see][for a review of this code]{ck-pca02}. The magnetograph analysis, based on the weak-field approximation, gives us a first approach of the vector magnetic field. Furthermore, these results were compared with more accurately ME and PCA-based inversions. \citet{ck-kuck09} concluded that the magnetic field strengths inferred with the magnetograph analysis are under-estimated by around $100 - 150$\,G, while the inclinations and azimuths are in very good agreement. 

The time series were inverted using the ME and the PCA-based inversion codes. \citet{ck-kuck09} presented, for both inversion codes, the best fits for the Stokes profiles of two selected points of the data-set and found that both techniques yielded almost the same magnetic field strengths, inclinations and azimuths. The Stokes profiles have Zeeman-like shapes and atomic polarization is almost absent at the filament.   

This paper will focus on the results achieved with the ME inversion code for the entire map of the AR filament.

\subsection{Milne-Eddington inversions}\label{kuckein-sec:MEinversion}
The \ion{He}{I} Stokes profiles were inverted using MELANIE \citep{ck-socas01}.
This inversion code computes the Zeeman-induced Stokes spectra, in the
incomplete Paschen-Back effect regime \citep[][]{ck-socas04}, that emerge from
a model atmosphere described by the Milne-Eddington approximation. A set of
eleven parameters of the model atmosphere are modified by the inversion code to
obtain the best fits to the observed Stokes profiles. We performed several
inversions for the same point changing the initial model atmosphere
configuration in order to obtain the best fit.  

\begin{figure*} 
\includegraphics[width=0.48\hsize]{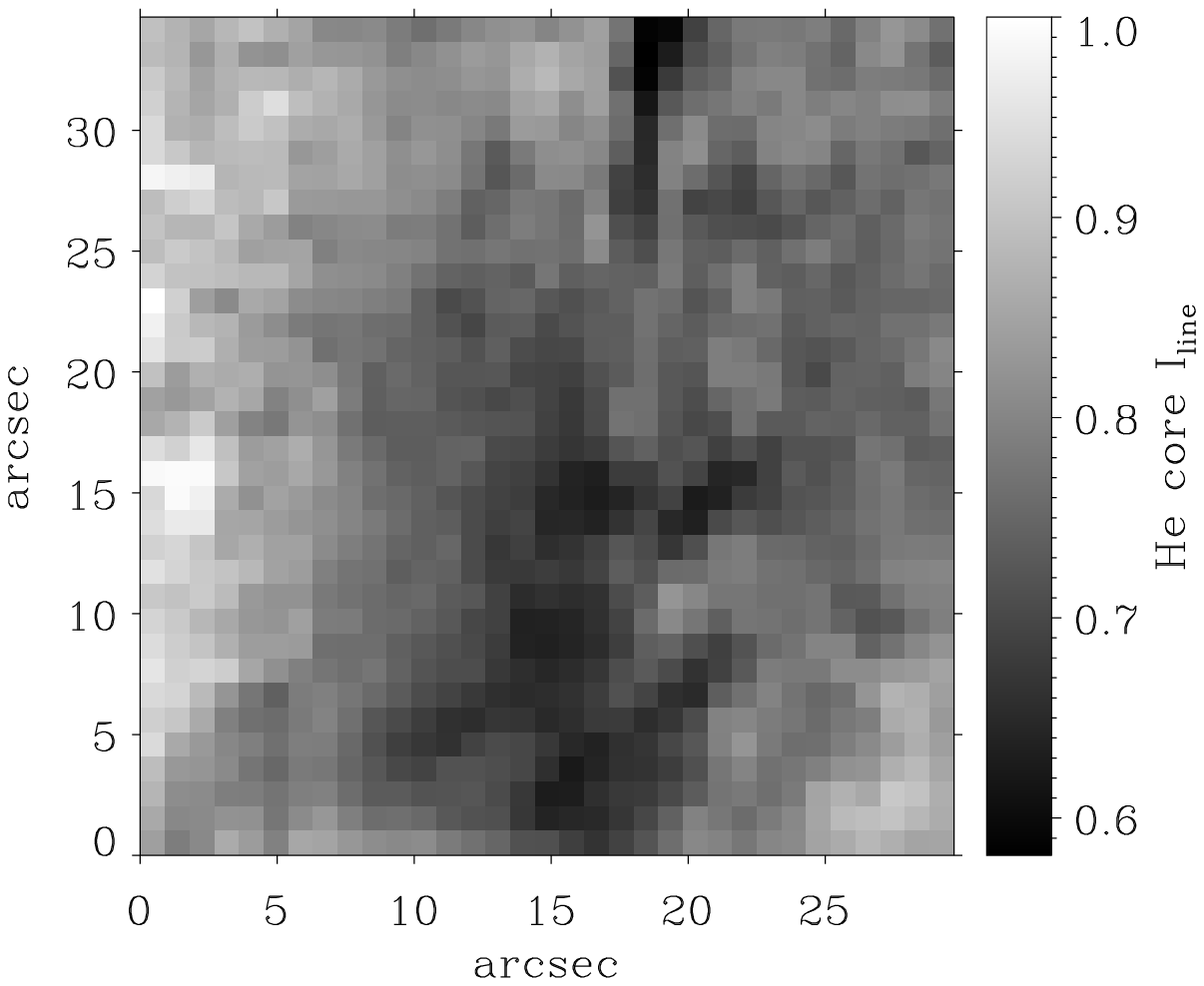}
\includegraphics[width=0.48\hsize]{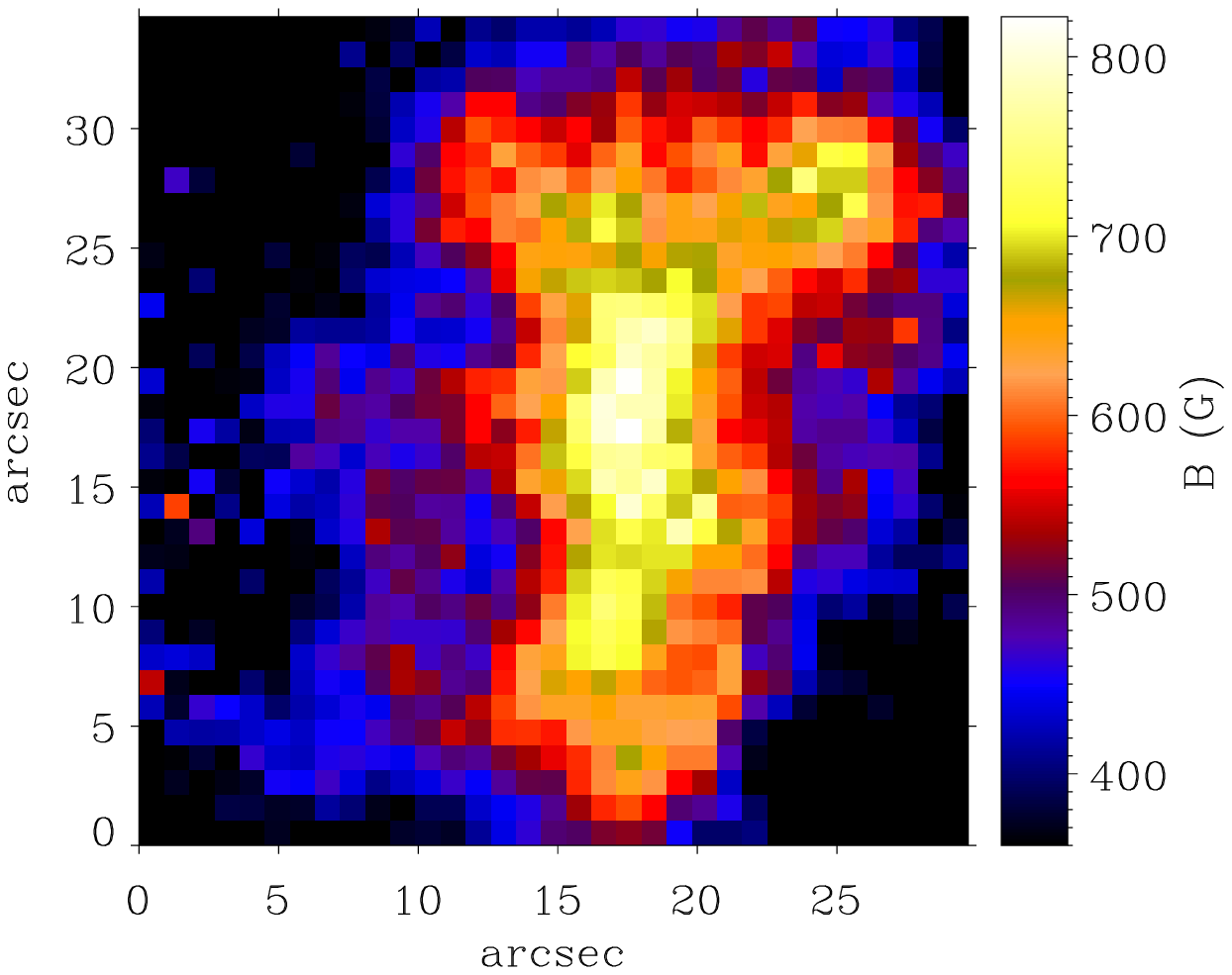} \\
\includegraphics[width=0.48\hsize]{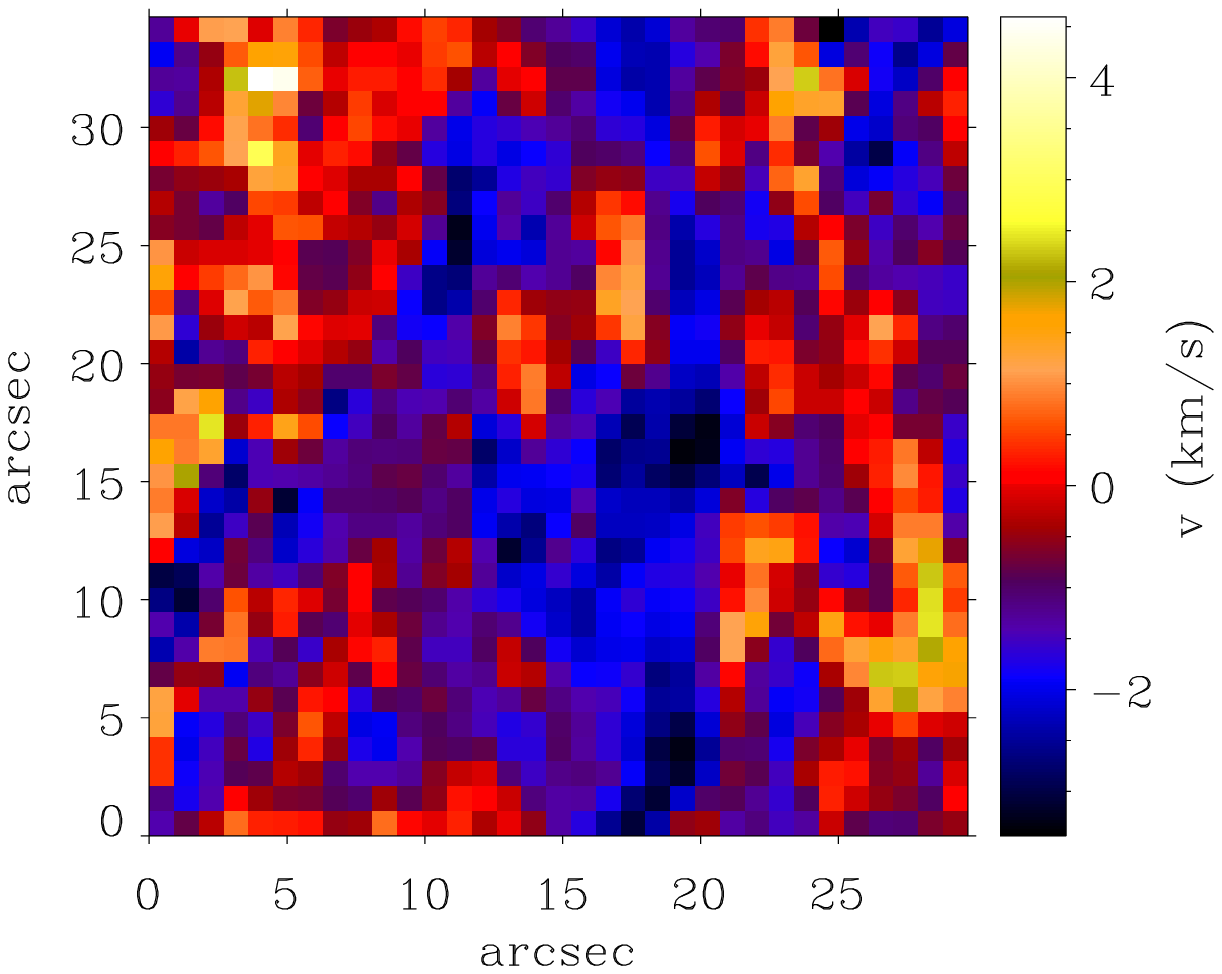}
\includegraphics[width=0.48\hsize]{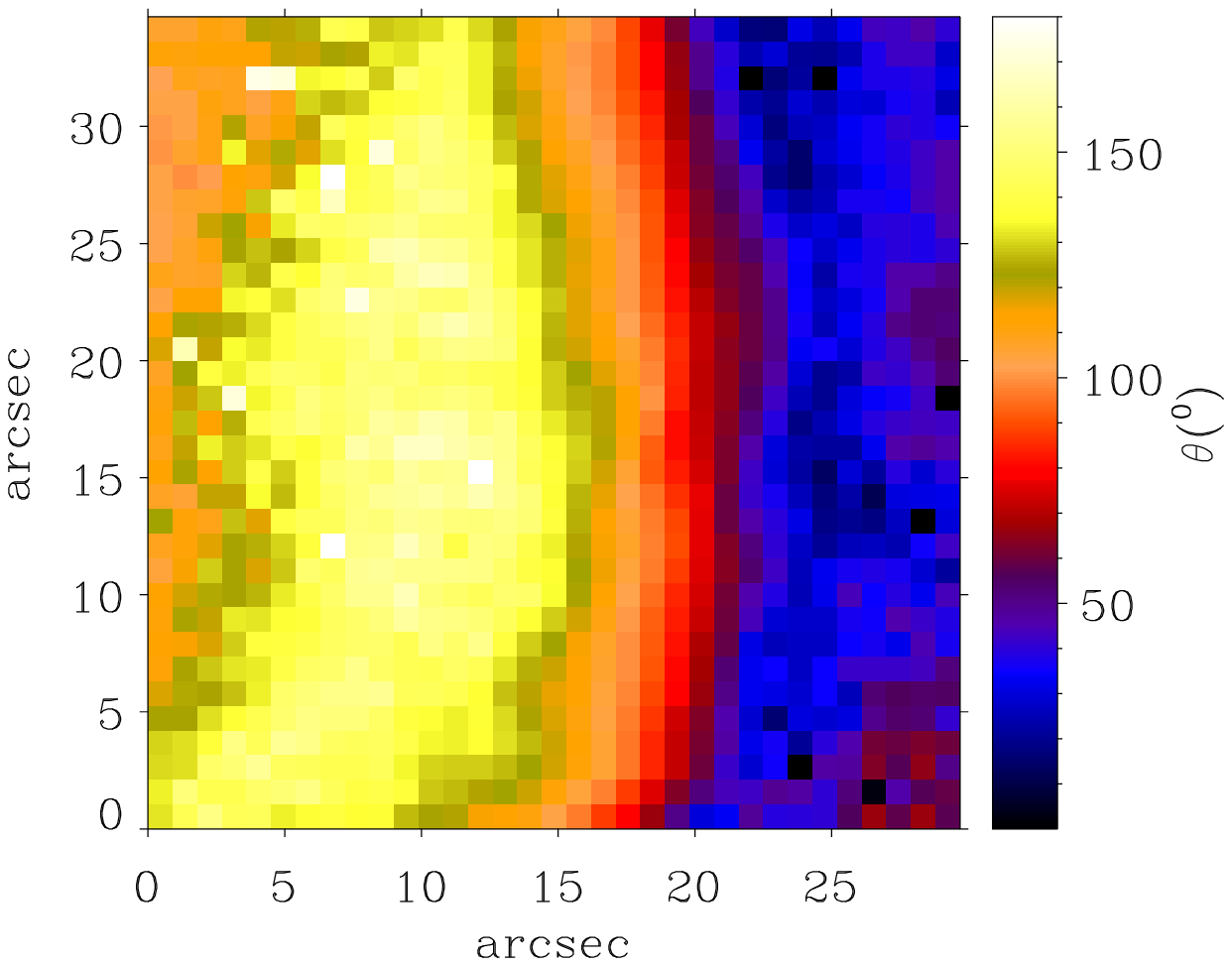} \\
\includegraphics[width=0.48\hsize]{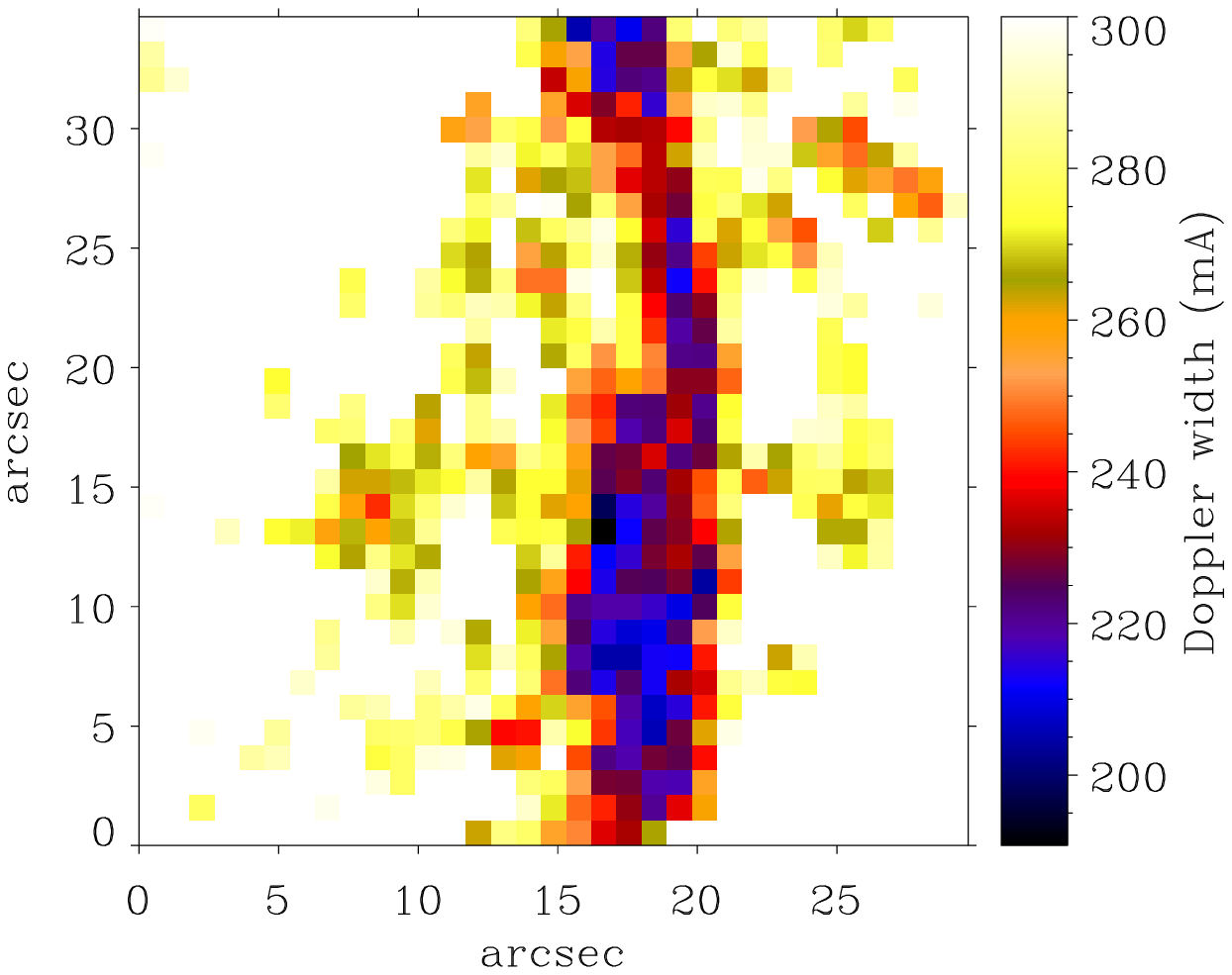}
\includegraphics[width=0.48\hsize]{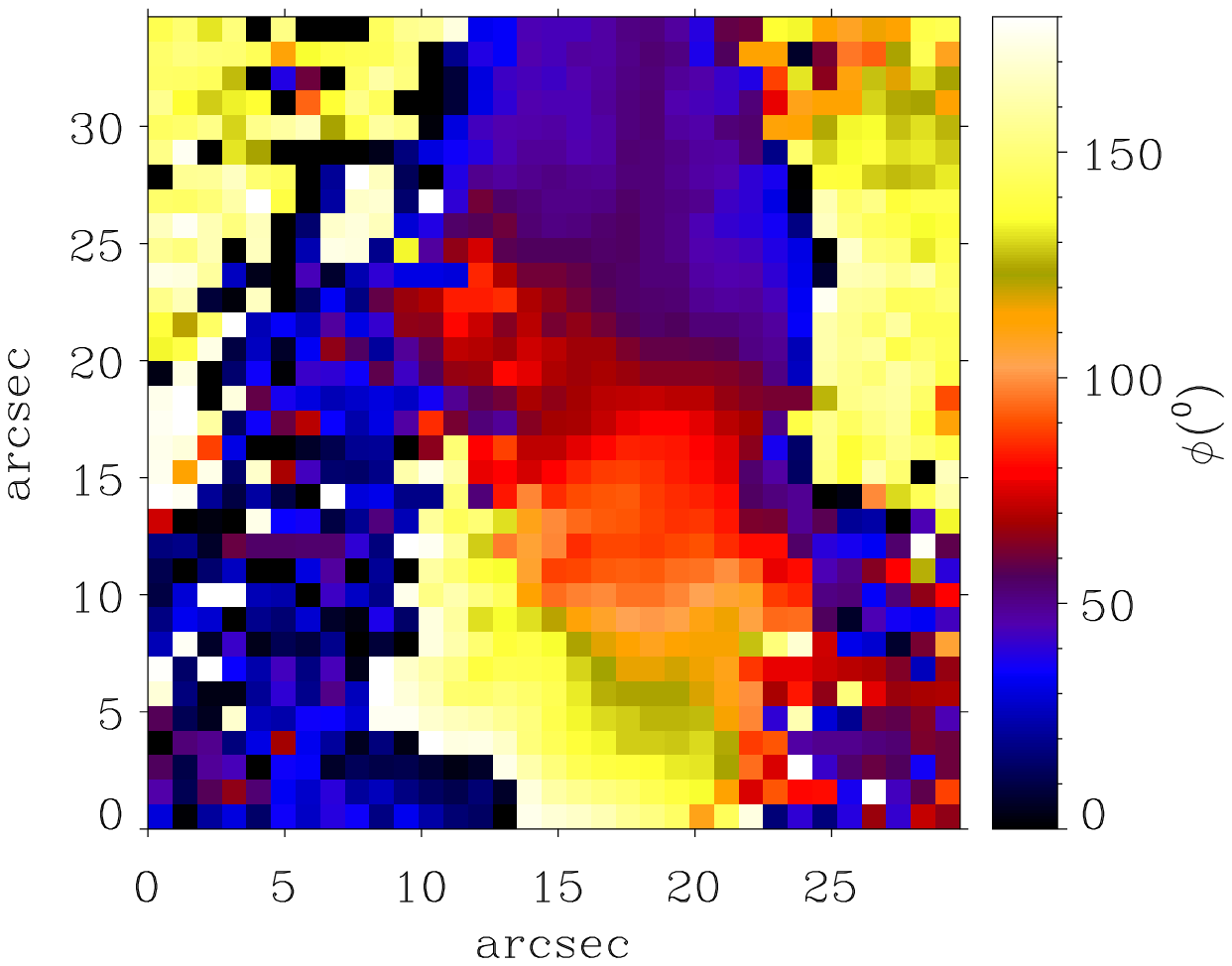}\\
\caption{\footnotesize From \textit{top} to \textit{bottom} and \textit{left} to \textit{right}: \ion{He}{I} red core intensity, magnetic field strength in Gauss units, LOS velocity (see definition in section \ref{kuckein-sec:MEinversion}), LOS inclination $\theta$ in degrees, Doppler width in m\AA\ and LOS azimuth $\phi$ in degrees.}  
\label{kuckein-fig:maps}
\end{figure*}

Figure \ref{kuckein-fig:maps} presents the magnetic field ($B$), LOS velocity
($v$), inclination ($\theta$) and azimuth ($\phi$) with respect to the
line-of-sight and the Doppler width obtained from the ME inversions. A
\ion{He}{I}, centered at the red core, slit reconstructed image is also shown
to identify the filament. On average, the magnetic field strength inferred from
the ME inversions in the filament is around $600 - 800$\,G. Moreover, high
inclination values between $80^\circ$ and $100^\circ$ with respect to the
line-of-sight (LOS) are found at the PIL indicating predominantly transverse
fields. 

The azimuth origin is referenced to the Earth's North-South direction. From the
lower right map of Fig. \ref{kuckein-fig:maps} it appears that the azimuth is
aligned with the dark structures of the filament. Nevertheless, these results
must be interpreted with caution because the 180 degree ambiguity is not yet
resolved. This is an important issue for future work.  

As can be seen from Figure \ref{kuckein-fig:maps}, the Doppler width is around
$210 - 220$\,m\AA\ in the filament, smaller than its surroundings and it seems to
shape the surface form of the filament. 

The LOS velocities were not calibrated for the orbital motions which contribute to the wavelength
shifts. Instead, we calculated the mean velocity from different areas of the map outside the filament and subtracted it from the velocity of every pixel. Consequently, this map gives as some hint
whether the plasma is moving upward (negative $v$) or downward (positive $v$). If we only take into account the velocities which correspond to Doppler widths below 220\,m\AA, the filament is rising with respect to its surroundings at mean speeds as large as $2.2$ km s$^{-1}$.

\section{Discussion}
We have presented Milne-Eddington inversions using the MELANIE code for an active region filament lying near the solar disk center. The Stokes profiles in the filament are completely dominated by the Zeeman effect. It is somewhat surprising that no atomic polarization was found. \citet{ck-casini09} suggest that the presence of an unresolved magnetic field with a random component could explain the lack of atomic polarization signatures in the Stokes profiles. Another explanation was proposed by \citet{ck-trujillo07}, who argue that the radiation field of the optical thick filament itself decreases the anisotropy radiation making it negligible. 

At the formation height of the \ion{He}{I} strong transverse magnetic fields are found of $600 - 800$\,G. In reviewing the literature, apart from our study, no observational evidences at chromospheric levels for such strong magnetic fields in AR filaments have been found yet. Is this common to all active region filaments? Further spectropolarimetric observations, with high resolution, are needed to answer this question. Recent photospheric observations of an AR filament, like the ones presented by \citet{ck-okamoto09}, showed horizontal magnetic fields of $650$\,G on average and evidence of an emerging helical flux rope. If our filament would lay at a very low height then we could expect such strong fields.

The inclination and azimuth maps have shown that the magnetic field vector is highly transversal and in some parts aligned with the dark structures of the filament. This alignment in the lower part of the AR would indicate that the field is inclined $\sim 45^{\circ}$ with respect to the axis of the filament. Since the velocity map of Fig. \ref{kuckein-fig:maps} shows upward motions for the filament, the emerging flux rope scenario \citep[see][]{ck-lites09} might be a valid explanation for this case. Furthermore, \citet{ck-lites09} proposes that the flux rope lies at low heights and therefore should become measurable using photospheric spectral lines. Since our spectral window comprises the photospheric \ion{Si}{I} $10827.1$~\AA\ line, it is mandatory to invert this line in order to retrieve information of its magnetic field vector in the photosphere.

 \begin{acknowledgements}
 Based on observations made with the VTT operated on the island of Tenerife by the KIS in the Spanish Observatorio del Teide of the Instituto de Astrof{\'i}sica de Canarias. Christoph Kuckein would like to thank the LOC of the NSO Workshop \#25 on \textit{Chromospheric Structure and Dynamics} for the travel support. The National Center for Atmospheric Research is sponsored by the National Science Foundation.
 \end{acknowledgements}

\bibliographystyle{aa}
\bibliography{kuckein}

\end{document}